\documentclass[aps,prc,twocolumn,floatfix,showpacs,amsmath,amssymb]{revtex4}
\usepackage{amssymb}
\usepackage{graphicx}

\begin{document}

\title{Neutron Transfer Reactions for Deformed Nuclei Using Sturmian Basis}
\author{
V. G. Gueorguiev$^{1}$, P. D. Kunz$^{2}$, J. E. Escher$^{1}$, F. S. Dietrich$^{1}$ \\
$^{1}$Lawrence Livermore National Laboratory, Livermore, California, USA\\
$^{2}$Department of Physics and Astrophysics, University of Colorado, Boulder, Colorado, USA
}\date{\today, ~ UCRL-JRNL-231626}

\begin{abstract}
We study the spin-parity distribution P(J$^{\pi}$,E) of $^{156}$Gd excited states 
above the neutron separation energy that are expected to be populated via the 
neutron pickup reaction $^{157}$Gd($^{3}$He,$^{4}$He)$^{156}$Gd.
In general, modeling of the spin-parity distribution is important for the applicability 
of the surrogate reaction technique as a method of deducing reaction cross sections.
We model excited states in $^{156}$Gd as rotational states built on intrinsic states 
consisting of a hole in the core where the hole represents neutron removal form a deformed 
single particle state. The reaction cross section to each excited state is calculated using 
standard reaction code that uses spherical reaction form-factor input. The spectroscopic 
factor associated with each form-factor is the expansion coefficient of the deformed neutron 
state in a spherical Sturmian basis consisting of the spherical reaction form-factors.
\end{abstract}

\maketitle


\section{Introduction and motivation}

Most light nuclei up to C and O are produced via nuclear reactions within the stars. 
Heavier nuclei near the valley of stability are produced via slow neutron capture process (s process).
To account for the observed abundance of the elements, however, one needs to also consider other 
processes such as proton capture process (p process) and rapid neutron capture process (r process).
The r process is essential in the production of the heaviest neutron rich elements since this process
generates nuclei far from the valley of stability that decay back towards the valley. Successively 
heavier neutron rich nuclei are produced through violent processes such as supernova explosions. 
Such nuclei usually  decay quickly. As a consequence, the understanding of the r-process contribution to the observed abundance of the elements requires knowledge of the neutron induced reactions on unstable nuclei. 
Unfortunately,  measuring these cross sections in a laboratory environment is a very difficult, if not 
impossible, task because of the technical and practical problems associated with the use of an unstable nuclei.

\begin{figure}[tbh]
\centerline {\includegraphics[width=8.5cm]{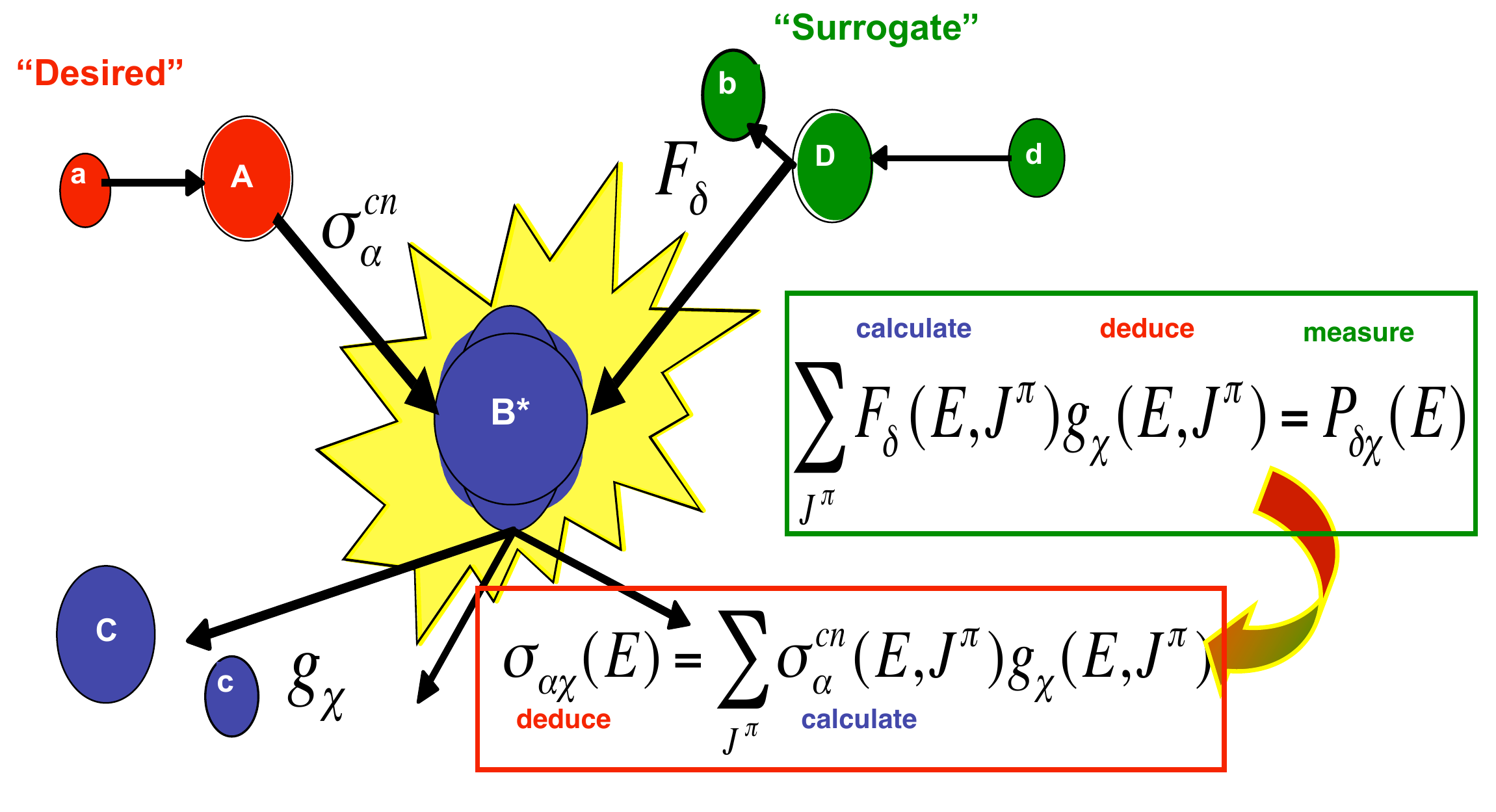}}
\caption{Reaction cross section for the desired reaction a+A $\rightarrow B^{\star}\rightarrow$ C+c 
might  involve an unstable target A. Alternatively, the cross section may be deduced using theoretical 
modeling and experimental data from a surrogate reaction d+D $\rightarrow b+B^{\star} \rightarrow$ C+c on 
a stable target D. Both reactions proceed via the formation of the same intermediate compound system B$^{\star}$.}
\label{TheSurrogateMethod}
\end{figure}

There have been various proposals for circumventing the problems associated with the use of unstable nuclei 
and yet to gain information about the desired nuclear reaction. One such approach is the Surrogate Method shown 
schematically in Fig. \ref{TheSurrogateMethod}.

In this paper we study the reaction $^{3}$He+$^{157}$Gd $\rightarrow$ $^{4}$He+$^{156}$Gd$^{\star}$ 
and model the formation probability of various excited states of the $^{156}$Gd system within a direct 
reaction framework. This reaction is a testing ground of the surrogate method for 
the neutron capture  reaction $^{155}$Gd+n $\rightarrow$ $^{156}$Gd$^{\star}$. 
Thus, one has to study the surrogate reaction for 
excitation energies $E_{ex}$ of $^{156}$Gd that correspond to few MeV neutron absorption  in 
$^{155}$Gd ($0<E_{n}<2$); by energy balance one has:
$E_{ex}=M_{^{155}Gd}+M_{n}+E_{n}-M_{^{156}Gd}=S_{n}+E_{n}$ where M is the corresponding 
rest mass and $S_{n}=8.536$ MeV is the neutron separation energy for $^{156}$Gd 
\cite{156Gd_list_of_levels}. 
Therefore, we shall focus on excited states in $^{156}$Gd such that:
\begin{equation}
E^{^{156}Gd}_{ex} > 8.5 ~ MeV.
\label{Eex bound}
\end{equation}

\section{Mathematical framework}

Atomic nuclei exhibit a multitude of spectral phenomena. Among the simplest
and relatively well understood properties are the rotational and vibrational
spectra as well as the general single-particle shell structure that explains
the observed magic numbers \cite{Bohr&Mottelson-I,Bohr&Mottelson-II}. The
traditional modeling of deformed even-even nuclei considers the rotational
states of nuclei as a collective phenomenon whose states are simple
rotational functions \cite{Bohr&Mottelson-II}. An extra particle is then
assumed to occupy a valence single particle state within a deformed
mean-field potential. This allows one to treat the odd-even deformed nuclei 
within the rotor plus single particle model. Within this
framework one can calculate cross sections for particle transfer as
transition amplitudes between collective rotational states and a particle plus
rotor states.

\subsection{States of axially deformed nuclei}
\label{SoADN}

\subsubsection{Bohr-Mottelson rotor model for deformed nuclei}

Within the rotor model an intrinsic state of a deformed system, 
with axial symmetry, gives rise to a rotational band with energies 
and wave functions of the form \cite{Bohr&Mottelson-II}: 
\begin{eqnarray}
J&=&K,K+(1+\delta_{K,0}),K+2(1+\delta_{K,0}), ...\nonumber\\
E(J)&=&E_{K}+\frac{\hbar^{2}}{2\cal{I}}(J(J+1)-K(K+1)),\\
\Psi_{KJM}&=&\frac{1}{4\pi}\sqrt{\frac{J(J+1)}{(1+\delta_{K,0})}}\times\nonumber\\
&\times&\left[ D_{MK}^{J}(\omega)\Phi _{K}
+(-1)^{K+J}D_{M-K}^{J}(\omega )\Phi _{\bar{K}}\right].\nonumber
\end{eqnarray}
Here $\cal{I}$ is the moment of inertia of the system, which for simplicity is assumed to be 
independent of $K$, $\omega $ stands for the angles $\theta $ and 
$\phi $, $D_{MK}^{J}$ are the rotational matrices, $\Phi _{\bar{K}}$ 
is the reflection of the intrinsic state $\Phi _{K}$ with respect to the 
symmetry axis, and $K$ is the angular momentum projection onto 
the symmetry axis.

\subsubsection{Rotor plus particle/hole system}

In the zero order approximation where, we neglect possible particle-core couplings
such as Coriolis coupling, the combined system of a particle/hole plus a core has an 
intrinsic state $\Phi _{\Omega}$ that can be viewed as a direct product of the  
intrinsic state of the core $\Phi^{cor}_{K}$ and the single particle/hole state $\psi_{\nu}$:
\begin{eqnarray}
\Phi _{\Omega}=\psi_{\nu}\Phi^{cor}_{K},~
E_{\Omega}=E_{K}+\epsilon_{\nu}.
\label{EOmEKEnu}
\end{eqnarray}
where  $\Omega=K+\nu$ due to axial symmetry. 
Therefore, for the particle plus core system we have:
\begin{eqnarray}
J&=&\Omega,\Omega+(1+\delta_{\Omega,0}),\Omega+2(1+\delta_{\Omega,0}), ...\nonumber\\
E(J)&=&E_{\Omega}+\frac{\hbar^{2}}{2\cal{I}}(J(J+1)-\Omega(\Omega+1)),\\
\Psi _{\Omega JM}&=&\frac{1}{4\pi}\sqrt{\frac{2J+1}{(1+\delta_{\Omega,0})}}\times\nonumber\\
&\times&\left[D_{M \Omega}^{J}(\omega)\Phi_{K}\psi_{\nu}
+(-1)^{\Omega+J}D_{M-\Omega}^{J}(\omega )\Phi _{\bar K}\psi_{\bar\nu}
\right].\nonumber
\end{eqnarray}

As long as the final states that we are interested in are assumed to be 
built out of the same initial intrinsic core state, we can make perturbation adjustments,
presumably small, to the single particle energies to incorporate BCS pairing and 
Coriolis coupling effects.

\subsection{Single particle states in axially deformed nuclei}

Since the nucleon density in nuclei is constant, except near the nuclear
surface, one expects that the general form of the mean-field potential,
which defines the single particle states, is of Woods-Saxon type \cite{Woods-Saxon:1954}: 
\begin{equation}
V(r)=\frac{V_{0}}{1+\exp \left( \frac{r-R}{a}\right) }  \label{Woods-Saxon}
\end{equation}
For spherical nuclei $R$ is constant and represents the position of the
nuclear surface while $a$ is related to the diffuseness of the potential near
the surface. For deformed nuclei, however, $R$ depends on the surface point
of interest and is often parametrized using spherical harmonics: 
\begin{equation}
R(\theta ,\phi )=R_{0}\left( 1+\sum_{\lambda \mu }a^{\lambda \mu }Y_{\lambda
\mu }\left( \theta ,\phi \right) \right)   \label{R in Multipoles}
\end{equation}
For axially deformed nuclei $R$ does not depend on $\phi $ due to the axial symmetry. 
The $\lambda=1$ is absent due to center of mass considerations and odd $\lambda$ terms 
are usually absent due to parity considerations; for axially deformed nuclei one usually
assumes quadrupole and hexapole deformation only: 
\begin{equation}
R(\theta ,\phi )=R_{0}\left( 1+\beta _{2}Y_{20}\left( \theta ,\phi \right)
+\beta _{4}Y_{40}\left( \theta ,\phi \right) \right)   \label{R in bettas}
\end{equation}

\subsubsection{Single particle states at small deformation}

For small values of the deformation parameters $\beta \lesssim 0.3$ in 
(\ref{R in bettas}) one can expand the Woods-Saxon potential (\ref{Woods-Saxon})
in Taylor series. The simplest approach is to consider only the first order
terms in the expansion \cite{Hird&Huang:CPC1975}. 
\begin{equation}
V(r,R)\simeq V(r,R_{0})-V^{\prime }(r,R_{0})\frac{R_{0}}{a}
(\beta_{2}Y_{20}\left( \theta ,\phi \right) +...)  \label{Linear WS approximation}
\end{equation}
While in many cases this seems to be sufficient, some particular single particle 
states are sensitive to small deformation values of beta $\beta_{2}\lesssim 0.1$. 

\subsubsection{Single particle levels at strong deformation}

If $\beta $ is sufficiently small, so that the Taylor expansion converges,
one can re-express $V(r,R)$ in multiple expansion in terms of spherical harmonics
\cite{Glendenning:2004}. An alternative way is to solve numerically the
Schr\"{o}dinger equation for the deformed Woods-Saxon potential 
\cite{Dudek&Nazarewicz}. When comparing the single particle energies calculated
numerically to the one calculated using only first order Taylor expansion
approximation, one finds that for rare-earth nuclei (nuclei near Gd) the
``Nilsson diagrams'' agree for $\beta _{2}\lesssim 0.1$ but start to deviate 
at larger deformations; in particular, there is a substantial deviation for 
$m_{j}=j$ states. 
\begin{figure}[tbh]
\centerline {\includegraphics[width=9.5cm]{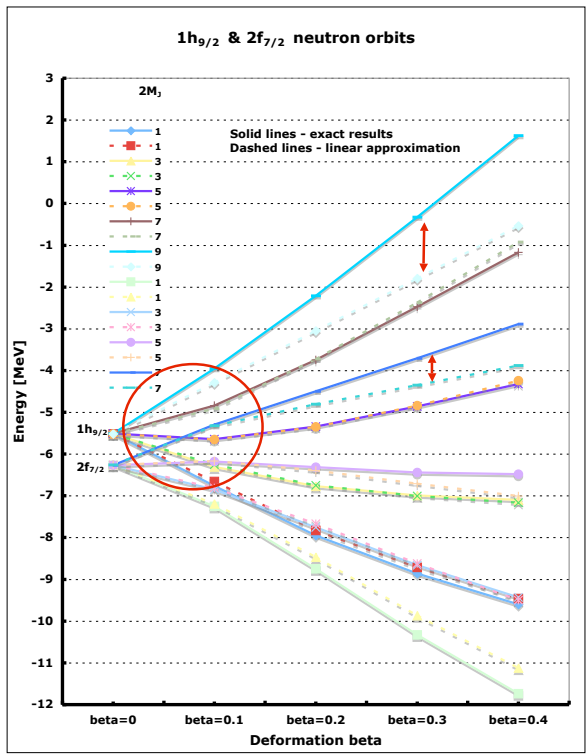}}
\caption{$^{157}$Gd deformed single particle energies as function of 
the deformation $\beta$ for $V_{0}$=-44.6883 MeV, $r_{0}$=1.27 fm, 
$a_{0}$=0.67 fm, $V_{ls}$=15.9945 MeV. The circle indicates the region of 
agreement between the linear treatment and the  exact numerical treatment. 
The arrows at $\beta=0.3$ indicate the most substantial deviation, between the 
linear treatment and the exact numerical treatment, observed for the $j=m=9/2$ 
and $j=m=7/2$ states.}
\label{spe Nilsson like diagram}
\end{figure}

To understand the impact of the deformation and what should be considered 
as a small deformation, calculations were performed using several codes
\cite{Dudek&Nazarewicz,Hird1973,Hird&Huang:CPC1975,TriaxialCode,
Rost:1967,DWUCK, CHUCK}.
First we made sure that in the case of zero deformation the low energy 
spectrum agrees well within few keV. In most cases an initial disagreement was fixed 
by improving the treatment of the reduced mass of the particle-core system. We also found 
that the choice of the spin-orbit interaction could result in a factor of about 2 or 4 
depending on the convention used. Usually obtaining the $p_{1/2}$ and $p_{3/2}$ 
splitting is the main indication that the factor affecting the strength of the 
spin-orbit term is properly accounted for. The energy of the $s_{1/2}$ state was 
used for understanding the convention for the strength of the central potential.
When the energies of the $s_{1/2}, p_{1/2}$, and $p_{3/2}$ states are reproduced, 
by each of the codes and the reduced mass is properly accounted for, the rest of 
the bound spectrum is usually well reproduced within few keV. 
For small non-zero deformation we used fewer codes 
\cite{Dudek&Nazarewicz,Hird&Huang:CPC1975,TriaxialCode,Rost:1967},
while for large deformation we relied only on \cite{Dudek&Nazarewicz}.

Fig \ref{spe Nilsson like diagram} shows the dependence of the single 
particle energies of the $nl_{j}=1h_{9/2}$ and $2f_{7/2}$ orbitals.
Calculations using a code that treats the interaction as linear in 
the deformation (\ref{Linear WS approximation})
are compared to an exact numerical treatment of the interaction 
\cite{Hird&Huang:CPC1975,Dudek&Nazarewicz}. 
Fig \ref{spe Nilsson like diagram} illustrates 
that one has to treat deformation with caution.

In Table \ref{DSPS} are shown the neutron bound states in $^{157}$Gd calculated 
using deformed Woods-Saxon potential \cite{Hird&Huang:CJP1975}. 
Since we have used WSBETA code \cite{Dudek&Nazarewicz}, which uses an exact 
numerical method instead of a linear approximation, we have set $\beta_{4}=0$ which
gives the binding energy of the 47$^{th}$ neutron single particle state close to 
the experimental neutron separation energy ($S_{n}=6.3598$ MeV \cite{157Gd_list_of_levels})  
in $^{157}$Gd $\epsilon_{47}=-6.361$ MeV.
 
\begin{table}[h]
{\center
\begin{tabular}{|c|cc|c|cc|c|cc|}
\hline
$\nu$ & $\epsilon_{\nu}$ [MeV] & $J^{\pi}$ & $\nu$ & $\epsilon_{\nu}$  [MeV] & $J^{\pi}$ & $\nu$ & $\epsilon_{\nu}$  [MeV] & $J^{\pi}$\\
\hline
1 &-39.971 &$\frac{1}{2}^+$ &23 &-16.900 &$\frac{1}{2}^-$ &45 &-7.054 &$\frac{11}{2}^+$\\
2 &-36.357 &$\frac{1}{2}^-$ &24 &-16.814 &$\frac{1}{2}^+$ &46 &-6.820 &$\frac{3}{2}^+$\\
\hline 
3 &-34.665 &$\frac{3}{2}^-$ &25 &-16.809 &$\frac{7}{2}^+$ &47 &{\bf -6.361} &$\frac{3}{2}^-$\\
4 &-34.036 &$\frac{1}{2}^-$ &26 &{\bf -14.870} &$\frac{1}{2}^+$ &48 &-5.980 &$\frac{5}{2}^+$\\
\hline
5 &-31.655 &$\frac{1}{2}^+$ &27 &-14.867 &$\frac{3}{2}^+$ &49 &-5.563 &$\frac{5}{2}^-$\\
6 &-30.453 &$\frac{3}{2}^+$ &28 &-14.661 &$\frac{9}{2}^+$ &50 &-5.000 &$\frac{1}{2}^-$\\
7 &-29.441 &$\frac{1}{2}^+$ &29 &-13.953 &$\frac{1}{2}^-$ &51 &-4.795 &$\frac{7}{2}^+$\\
8 &-28.570 &$\frac{5}{2}^+$ &30 &-13.408 &$\frac{3}{2}^-$ &52 &-4.290 &$\frac{5}{2}^-$\\
9 &-26.745 &$\frac{3}{2}^+$ &31 &-12.608 &$\frac{3}{2}^+$ &53 &-3.284 &$\frac{9}{2}^+$\\
10 &-26.230 &$\frac{1}{2}^-$ &32 &-12.424 &$\frac{5}{2}^-$ &54 &-2.846 &$\frac{7}{2}^-$\\
11 &-26.045 &$\frac{1}{2}^+$ &33 &-12.286 &$\frac{5}{2}^+$ &55 &-2.706 &$\frac{1}{2}^+$\\
12 &-25.345 &$\frac{3}{2}^-$ &34 &-11.239 &$\frac{1}{2}^+$ &56 &-2.655 &$\frac{1}{2}^-$\\
13 &-23.895 &$\frac{5}{2}^-$ &35 &-11.051 &$\frac{7}{2}^-$ &57 &-2.072 &$\frac{3}{2}^-$\\
14 &-23.609 &$\frac{1}{2}^-$ &36 &-9.986 &$\frac{5}{2}^+$ &58 &-1.951 &$\frac{7}{2}^-$\\
15 &-21.863 &$\frac{7}{2}^-$ &37 &-9.518 &$\frac{1}{2}^-$ &59 &-1.427 &$\frac{11}{2}^+$\\
16 &-21.349 &$\frac{3}{2}^-$ &38 &-9.290 &$\frac{9}{2}^-$ &60 &-1.296 &$\frac{3}{2}^+$\\
17 &-20.911 &$\frac{1}{2}^-$ &39 &-8.994 &$\frac{7}{2}^+$ &61 &-0.820 &$\frac{1}{2}^+$\\
18 &-20.294 &$\frac{1}{2}^+$ &40 &-8.137 &$\frac{1}{2}^-$ &62 &-0.716 &$\frac{3}{2}^-$\\
19 &-19.616 &$\frac{3}{2}^+$ &41 &-8.053 &$\frac{1}{2}^+$ &63 &-0.390 &$\frac{1}{2}^-$\\
20 &-18.437 &$\frac{5}{2}^+$ &42 &-7.811 &$\frac{3}{2}^-$ &64 &-0.131 &$\frac{1}{2}^-$\\
21 &-18.321 &$\frac{5}{2}^-$ &43 &-7.591 &$\frac{3}{2}^+$ &65 & 0.015 &$\frac{3}{2}^-$\\
22 &-18.009 &$\frac{3}{2}^-$ &44 &-7.276 &$\frac{1}{2}^+$ &66 & 0.400 &$\frac{5}{2}^+$\\
\hline
\end{tabular}
}
\caption{Neutron bound states in $^{157}Gd$ calculated with the WSBETA code 
\cite{Dudek&Nazarewicz} using Woods-Saxon parameters from Ref. \cite{Hird&Huang:CJP1975}
but $\beta_{4}=0$ so that the 47$^{th}$ neutron state ($\epsilon_{47}=-6.361$) is near 
the experimental neutron separation energy $S_{n}=6.3598$ MeV \cite{157Gd_list_of_levels},
$V_{0}$=-45.1776 MeV, $r_{0}$=1.25 fm, $a_{0}$=0.65 fm, $V_{ls}$=19.2015 MeV, 
$\beta_{2}=0.29$ and $\beta_{4}=0$. The Fermi level and the level that results in excitation 
near the neutron separation energy in $^{156}$Gd
(S$_{n}$= 8.536 MeV) are in boldface and marked by two horizontal lines.}
\label{DSPS}
\end{table}
 
\subsection{Adjustments to the single particle energies}

In the Distorted Wave Born Approximation (DWBA) the single particle wave function of the transferred particle is of essence along with the probability amplitude of the particle to be in that state. Ideally one would calculate the many body eigenstates of the initial and final system within a chosen single particle basis and then would deduce the probability amplitude - the spectroscopic factor for the single particle to be in any of the single particle states. Such approach has  the advantage that a many body Hamiltonian that describes the initial and the final systems is used to deduce the single particle wave function of the transferred particle. However, finding the many body Hamiltonian that works equally well across the nuclear chart has proven elusive. 

The absence of clearly successful and unique many body Hamiltonian and method for solving the many body problem justifies the use of simpler models for deducing the single particle wave function of the transferred particle such as the particle plus core model described in section \ref{SoADN}. There are some issues with such models. For example, the total anti-symmetrization, pairing interaction, and particle-core interaction effects are points to be addressed. The effect of the anti-symmetrization is usually taken into account within the DWBA calculation by using the appropriate combinatorial factors
\cite{Satchler:1958}. The particle-core interaction is usually dominated by a Coriolis coupling and the pairing effects within the core are accounted for through the BCS pairing model. Other interaction such as coupling to the quadrupole vibrations of the core will not be considered here 
\cite{Azziz&Covello:1969}.

\subsubsection{BCS pairing}

The pairing interaction is a many body phenomenon which is effectively an interaction between pairs of time conjugated particles. Exact solutions of the pairing and its variations are well known \cite{PhysRev.144.874,dukelsky:072503,pan:112503}. However, an approximation know as BCS paring is very convenient \cite{papenbrock:014304}. Since the pairing interaction is effectively pair creation and annihilation  process its main effect is on the occupation of the single particle states near the Fermi energy that represents the location of the last occupied state and next unoccupied state. This means that the occupation number $n_{\nu}$ of states that are sufficiently far from the Fermi energy $\mu$ are practically not affected by the paring.
\begin{equation}
n_{\nu}=1-\frac{\epsilon_{\nu}-\mu}{\sqrt{(\epsilon_{\nu}-\mu)^{2}+\Delta^{2}}}
\label{n_nu}
\end{equation}

\begin{figure}[tbh]
\centerline {\includegraphics[width=8.5cm]{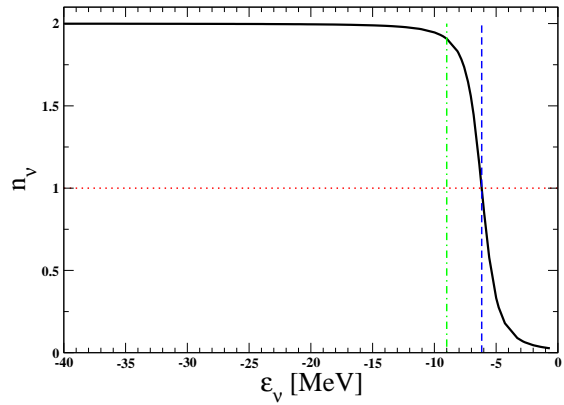}}
\caption{Occupation number $n_{\nu}$ for single particle bound states $\epsilon_{\nu}$ in 
the $^{157}$Gd system. Single particle energies $\epsilon_{\nu}$ are calculated for deformed Woods-Saxon potential with parameters: $V_{0}$=-45.1776 MeV, $r_{0}$=1.25 fm, $a_{0}$=0.65 fm, $V_{ls}$=19.2015 MeV, deformation $\beta_{2}=0.29$ and the following BCS parameters: Fermi energy $\mu=-6.169$ MeV, pairing gap $\Delta=1.307$ MeV, and pairing strength $g=0.15$ MeV.}
\label{BCSPairing}
\end{figure}

Fig. \ref{BCSPairing} shows the occupation number $n_{\nu}$ by applying the BCS equations to the set of single particle energies shown in Table \ref{DSPS}. From the graph one can see that the occupation of the particle levels with energies $\epsilon_{\nu}<-9$ MeV are practicly unaffected by the pairing interaction. Notice that according to (\ref{Eex bound}) one has to pull out neutrons that are at least 8 MeV below the Fermi level: 
\begin{equation}
\epsilon_{\nu}< -14.9 ~MeV. 
\label{enu}
\end{equation}

For this BCS calculation, we used all neutron bound states in 
$^{157}$Gd except the 47$^{th}$ state which is assumed to be Pauli bocked to the pairing interaction due to the odd nucleon. The Fermi energy $\mu$ was chosen to be between the last occupied state and the first unoccupied state in the system $\mu=(\epsilon_{47}+\epsilon_{48})/2$. The values of the pairing gap $\Delta$ and the pairing strength $g$ were determined from the BCS equations.

\subsubsection{Particle core coupling}

It is well know that the Coriolis coupling is an important interaction 
\cite{Kanestrom&Tjom:1969,Joshi&Sood:1974,Peng&Maher:1976,granja:034316}.
\begin{equation}
H_{C}=-\frac{\hbar^{2}}{2\cal{I}}(I_{+}j_{-}+I_{-}j_{+})
\label{Coriolis coupling}
\end{equation}
The proper consideration results in Coriolis band mixing and technical complications similar 
to the shell model wave function type approach. It is known that the first order perturbation to the energy is non-zero for  $\Omega=\frac{1}{2}$ bands only and higher order terms are needed to obtain the energy shifts in $\Omega>\frac{1}{2}$ bands \cite{Kanestrom&Tjom:1969,Joshi&Sood:1974}.

For this study, however, we consider only one intrinsic state of the core (the ground state of the target) 
coupled to various neutron holes in the core. This, therefore, eliminates any Coriolis band mixing 
from our model.  

There is, however, an important particle/hole core coupling that we estimate by using first order 
perturbation approximation. This gives us the  energy splitting between the states $\Omega=|K-\nu|$ and 
$\Omega=K+\nu$. It is clear that according to the simple non-interaction particle plus rotor model (\ref{EOmEKEnu}), 
we have two degenerate states $\Omega=|K-\nu|$ and $\Omega=K+\nu$. To estimate the splitting of these two states, 
due to rotation, we look at the expectation value of the rotational energy of the particle plus core system in the 
intrinsic frame (\ref{EOmEKEnu}):
\begin{eqnarray*}
&&\left< \Omega=K\pm\nu \right| 
( \vec{J}_{cor}+\vec{j}_{p} )^{2}
\left| \Omega=K\pm\nu \right>=\\
=&&\left< K \right| \vec{J}_{cor}^{2} \left| K \right>+
\left< \nu \right| \vec{j}_{p}^{2} \left|\nu \right>+\\
&& + 2\left< \Omega=K\pm\nu \right|
\vec{J}_{cor}\cdot\vec{j}_{p}
\left| \Omega=K\pm\nu \right>=\\
=&&\left< K \right| \vec{J}_{cor}^{2} \left| K \right>+
\left< \nu \right| \vec{j}_{p}^{2} \left|\nu \right> \pm 2 K \nu.
\end{eqnarray*}
Since the first two terms are the same and represent the rotational energy 
of the core  and the particle, one expects the energy splitting to be:
\begin{equation}
\Delta E(\Omega_{\pm}=|K\pm\nu|)=\pm c\frac{\hbar^{2}}{2\cal{I}}K\nu.
\label{Eshift}
\end{equation}
Here $c$ is a coefficient that relates the moment of inertia around the symmetry axis $z$ to 
the moment of inertia around $x$ and $y$ axes.  One could estimate its value from the mathematical 
expressions for the  moment of inertia of a rigid ellipsoid as given in the Appendix section \ref{Omega_pm splitting}. 
Here, however, we prefer phenomenological adjustment that reproduces the experimentally observed 
splitting between $\Omega =0$ and $\Omega=3$ bands in $^{156}$Gd by using the experimental moment 
of inertia $\cal{I}$ needed to reproduce the $\Omega =0$ ground state band.

\subsection{Reaction cross sections for deformed nuclei}

\subsubsection{DWBA direct reactions within the Optical Model Potential (OMP)\ theory}

The Distorted Wave Born Approximation (DWBA) considers the initial (final) 
state of the system to be a product of initial $|A,a>$ (final $|B,b>$) intrinsic 
states  of the target, projectile, and distorted wave function $\chi^{\pm}$ 
that depends on the relative coordinates between the target and the projectile 
\cite{Satchler:1958,DWUCK, CHUCK}. The reaction cross section is then 
calculated using an Optical Model Potential (OMP) for $\chi^{\pm}$ and the 
appropriate nuclear interaction $V$ relating the initial state $|A,a>$ with the 
final state final $|B,b>$. The transition amplitude then depends on the 
corresponding interaction matrix $<Bb|V|Aa>$. For one-nucleon transfer 
reactions this matrix element depends on the initial and final state of the 
transferred nucleon, the particular interaction that connects these states, and 
the overlap of the other nucleons. For a neutron pickup reaction, it reduces to: 
the neutron wave function $\psi$ inside the target, the strength $D_{0}$ 
of a zero range $\delta$ force interaction with the projectile, and a spectroscopic 
factor $S^{1/2}$ showing how much the target $A$ looks like a remnant $B$ 
plus a nucleon in the state $\psi$  ($<A|\psi^{\dagger}B>$).

\subsubsection{Separation energy matching method for proper tail of the 
single spherical states}

For transfer reactions that result in low energy excited states the correct 
asymptotic tail of the wave function, which is related to
the neutron separation energy, is very important in neutron pickup since
one expects that the last (outer most) nucleon is being transferred. Using
wave functions that have the desired binding energy by adjusting the depth
of the Woods-Saxon binding potential is one of the simplest and usually very
successful approximations for calculating the reaction cross section. 
In this  approach one keeps the geometric factors of the binding potential, 
such as $r_{0}$ and $a_{0}$ of the potential, fixed from systematics but changes the 
depth of the potential until a bound state $\psi$ with the desired binding energy is found. 
The approach has even been extended to the point where the variable parameter is 
the strength of a surface derivative potential.

\subsubsection{Sturmian method for form-factors}

Problems with the separation energy matching using only one spherical wave
function has been a motivation for more accurate description of the relevant form 
factor by inclusion of more than one basis state. The spherical harmonic oscillator (SHO) 
is a traditional basis to expand a wave function. However, the tail in a finite SHO basis 
expansion is allays wrong. An interesting alternative basis to expand a deformed bound 
state is to use the set of bound states of the spherical part of the interaction 
\cite{Hird&Huang:CJP1975}. However, this expansion does not guarantee correct tail either. 
An expansion would provide the correct tail if one of the basis states is a bound state with 
the same energy as the state that is being expanded and all the other non zero components 
correspond to basis states with a deeper binding energy. 

In a Sturmian basis all the basis states have the same tail as the original state 
that is being expanded in this basis. In order to maintain correct asymptotic 
tail one is now facing a problem where one has to find 
different wave functions and potential strengths that result in the same energy and thus the 
same wave function tail.  The Sturmian basis method has been utilized before 
\cite{Andersen&Back:1970}. The method in \cite{Andersen&Back:1970} relies essentially on 
expressing the deformed potential as liner in $\beta_{2}$ which may not be sufficient in the case 
of strong deformation. Unfortunately, this approach relies on the linear form of the potential with 
respect to the deformation parameters.

Another, more successful and physically more appealing approach is to
consider coupled channel (CC) calculations \cite{Rost:1967}. Unfortunately, this code  
in its current version has only $\beta_{2}$ deformation. In one of the next section we 
will discuss briefly our comparison to this code and possible further developments. 

To illustrate the role of various parameters involved, we now look at the incoherent
DWBA reaction cross section for a particle transfer from a deformed 
single particle state $\psi_{\nu}$ that can be expressed in terms of transfer cross 
sections on spherical single particle states $\phi_{nlj}$
\cite{Satchler:1958,Andersen&Back:1970}:
\begin{equation}
d\sigma(J_{i}K_{i} \rightarrow J_{f}K_{f};\nu)=\sum_{lj}\sum_{n}(a_{\nu}v_{\nu}c_{\nu}^{nlj})^{2}
d\sigma^{DW}_{nlj}
\end{equation}
Here $\sigma^{DW}_{nlj}$ are the DWBA cross section from a spherical state $\phi_{nlj}$, 
$c_{\nu}^{nlj}$ are the expansion coefficients of the  state $\psi_{\nu}$ in 
spherical basis states $\phi_{nlj}$, $v_{\nu}$ represents BCS occupation number 
($v_{\nu}^{2}=n_{\nu}/2$) of the state $\nu$, and $a_{\nu}$ is the Coriolis band 
mixing amplitude. The spectroscopic factor $S_{lj}$ is often used as shorthand 
notation for the term raised to second power. In our calculations, we actually consider the 
generally more appropriate coherent cross section by using super-position of basis states 
$\phi_{nlj}$ with amplitudes $a_{\nu}v_{\nu}c_{\nu}^{nlj}$.

\section{Neutron transfer reaction results}

In this section we present the results of our calculations on the neutron transfer reaction $^{157}${Gd}($^{3}${He},$^{4}${He})$^{156}${Gd}. We first discuss the OMP for the in-going and 
out-going channels, then we show the excited states in $^{156}$Gd according to the particle 
plus rotor model which in this case is actually a hole in the core model. Next we briefly discus the model space convergence in the determination of the $c_{\nu}^{nlj}$ amplitudes. After that 
the absolute cross sections and their convergence are discussed along with a smeared 
$\sigma(E)$ cross section profile near the neutron separation energy. Finally, we turn to a 
discussion of the probability distributions $P(J^{\pi},E)$.

Our transfer cross sections related to the reaction $^{157}$Gd($^{3}$He,$^{4}$He)$^{156}$Gd 
are for one nucleon removal from a deformed single particle state $\psi_{\epsilon_{\nu}}$  from 
the $^{157}$Gd system viewed as a core in a $K=3/2^{-}$ state. 
Presumably, the final states of the $^{156}$Gd system are rotational states built on the
intrinsic state $\Omega$ just like in the particle plus core model (\ref{EOmEKEnu}).
However, instead particles we work with holes in the core 
$|A=156,{\Omega}>=\psi^{\dagger}_{\nu} ~|{A=157},K={3/2}^{-}>$.

\subsection{Elastic cross sections and the optical model parameters}

Usually, calculations with distorted waves within the DWBA use an Optical
Model Potential (OMP): 
\begin{eqnarray*}
U &=&Vf(x_{0})+iWf(x_{w})-V_{so}\frac{1}{r}\frac{df(x_{so})}{dr}\vec{l}\cdot 
\vec{s} \\
f(r) &=&\frac{1}{1+\exp \left( (r-R)/a\right) },\quad
R=r_{0}A^{1/3}
\end{eqnarray*}

\begin{figure}[htb]
\centerline {\includegraphics[width=8.5cm]{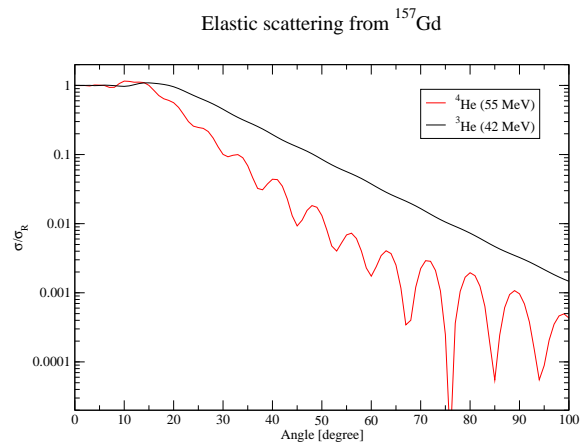}}
\caption{Rutherford normalized elastic cross sections for the relevant He+Gd
reaction. The smoothness of the  $^{3}$He cross section is due to the lower value 
of the associated energy.}
\label{He+157Gd_elastic_Normalized}
\end{figure}

Table \ref{DW OMP parameters} shows the OMP parameters used in our calculation. 
These parameters are taken from the Reference Input Parameter Library (RIPL-2 ) \cite{RIPL-2}.
The Rutherford normalized elastic cross sections for the ingoing and out-going channels are shown in  
Fig. \ref{He+157Gd_elastic_Normalized}. Notice that the absence of interference pattern for the 
$^{3}$He cross section is due to its relatively low energy.
\begin{table}[h]
{\center
\begin{tabular}{|c|c|c|c|c|c|c|c|c|c|c|}
\hline
Reaction & $r_{c}$ & V & $r_{0}$ & $a_{0}$ & W & $r_{w}$ & $a_{w}$ & $V_{so}$
& $r_{so}$ & $a_{so}$ \\ \hline
$^{3}$He + $^{157}$Gd & 1.3 & 154 & 1.2 & 0.72 & 36 & 1.4 & 0.88 & 2.5 & 1.2
& 0.72 \\ 
$^{4}$He + $^{156}$Gd & 1.2 & 159 & 1.2 & 0.8 & 14 & 1.6 & 0.7 & 0 & 1.2 & 
0.8 \\ \hline
\end{tabular}
}
\caption{OMP parameters, in fm and MeV units, for the in-going $^{3}$He+$^{157}$Gd 
\cite{Becchetti&Greenleesg:1969} and the out-going $^{4}$He+$^{156}$Gd 
\cite{Avrigeanu&Hodgson:1994}.}
\label{DW OMP parameters}
\end{table}

\subsection{Excited states in $^{156}$Gd according to the hole in the core model}

In calculating a transfer reaction to an excited state in $^{156}$Gd, we first construct an intrinsic state in 
$^{156}$Gd as a hole in the $^{157}$Gd core then we consider rotational states built on that intrinsic state. 
We consider total of 12 members of the related $\Omega=|K\pm\nu|$ rotational bands to include 
all possible excitations with $J<8$.
\begin{eqnarray}
|^{156}Gd,{\Omega=|K\pm\nu|}>&=&\psi^{\dagger}_{\pm\nu} |^{157}Gd,K={3/2}^{-}>\\
E(J^{\pi};\Omega=|K\pm\nu|)&=&\epsilon_{0}-\epsilon_{\nu}+\frac{\hbar^{2}}{2\cal{I}}
(J(J+1)+\delta_{\pm}) \nonumber
\end{eqnarray}
Here $\epsilon_{0}$ is used to set the ground state energy of $^{156}Gd$ to zero, and $\delta_{\pm}$
is the energy shift of the $\Omega=K+\nu$ state relative to the $\Omega=|K-\nu|$ state. Thus if we set 
$\delta_{-}=0$ for $\Omega=|K-\nu|$ then $\delta_{+}=2cK\nu$ for $\Omega=K+\nu$ states. We always keep 
$\Omega,\nu$, and $K$ positive when we use them as quantum numbers of the states.

By fitting the first four excited states of the $\Omega=0^{+}$ ground state band, the moment of inertia is fixed to be ${\hbar^{2}/2\cal{I}}=13.59$ KeV for the rotational bands in $^{156}$Gd. The $c=17.741$ coefficient for the energy shift (\ref{Eshift}) of the related $\Omega=3^{+}$ band was chosen to reproduced the excitation energy of the 
$3_{1}^{+}$ state. The value of $c$ is about 1.65 times the ratio of the 
moments of inertia $\cal{I_{||}/\cal{I_{\perp}}}$ for rigid ellipsoid with $\beta=0.29$.
In Table \ref{YrastBand} is shown comparison of the experimentally observed  low energy states in 
$^{156}$Gd \cite{157Gd_list_of_levels} and excitation energies calculated considering a neutron hole  coupled to a core.
\begin{table}[h]
{\center
\begin{tabular}{|c|c|c||c|c|c|}
\hline
$J_{n}^{\pi}$ & $E_{exp}$ [MeV] & $E_{th}$ [MeV]& $J_{n}^{\pi}$ & $E_{exp}$ [MeV] & $E_{th}$ [MeV]\\
\hline
0$_{gs}^{+}$ &  0 &  0.004 &  3$_{1}^{+}$ &  1.248 &  1.255 \\
\hline
2$_{1}^{+}$ &  0.089 &  0.085 &  4$_{3}^{+}$ &  1.355 &  1.363\\
4$_{1}^{+}$ &  0.288 &  0.275 &  5$_{1}^{+}$ &  1.507 &  1.499\\
6$_{1}^{+}$ &  0.585 &  0.574 &  6$_{3}^{+}$ &  1.644 &  1.662\\
8$_{1}^{+}$ &  0.965 &  0.982 &  7$_{1}^{+}$ &  1.85   &  1.852\\
\hline
\end{tabular}
}
\caption{Comparison of the experimentally observed  low energy states in $^{156}$Gd 
\cite{157Gd_list_of_levels} and excitation energies calculated considering a neutron hole 
$\nu=3/2^{-}$ at $\epsilon_{47}=-6.361$ coupled to the core system $^{157}$Gd with $K=3/2^{-}$.
The first three columns represent the $\Omega=0^{+}$ ground state band; the next three columns 
represent the $\Omega=3^{+}$ band. The theoretical ground state is not exactly zero because 
$\epsilon_{47}$ is not exactly equal the neutron separation energy $S_{n}$ in $^{157}$Gd.}
\label{YrastBand}
\end{table}

\begin{figure}[htb]
\centerline {\includegraphics[width=9.5cm]{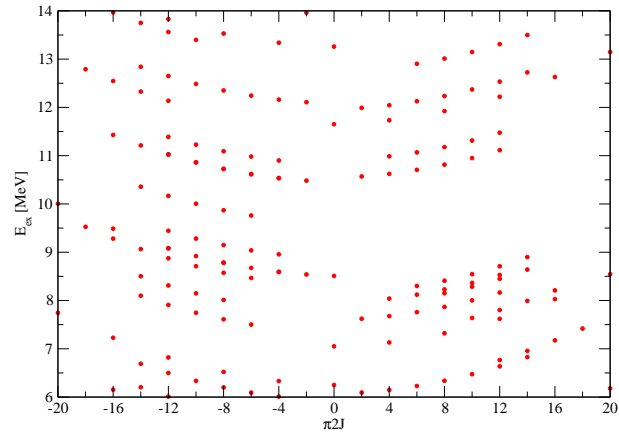}}
\caption{Distribution of the calculated $^{156}$Gd excited states. The sign of the x-axis 
encodes the parity of the states and the magnitude is equal to the twice of the angular momentum J.}
\label{pi2JE}
\end{figure}

In Fig. \ref{pi2JE} are shown the parity ($\pi$), angular momentum ($J$), and energy ($E$) of the states that are considered in the calculations of the reaction cross sections $\sigma(J^{\pi};E)$. The absence of positive parity states in the 10 MeV region is due to the five positive parity single particle states 24-28 with energy between -16.81 and -14.66  in Table \ref{DSPS}. These states will produce a parity asymmetry in the $P(J^{\pi},E)$ distribution in the energy region of interest for the surrogate method.

\subsection{Model space convergence}

In order to compute the transfer cross section one needs the single particle wave function and the corresponding spectroscopic factor. We already discussed our main approximations to the spectroscopic factor: we neglect the Coriolis band mixing ($a_{\nu}=1$), the pairing effects ($v_{\nu}=1$), and consider single particle wave functions that are expressed in a Sturmian basis. The idea is to maintain the same exponential tail for each spherical basis state as the tail of the deformed state that we are interested in. We have to make expansion in a spherical basis in order to be able to use existing reaction codes. 

Here we consider the Sturmian approach by using Sturmian spherical basis states. We calculate the spectra of a deformed Woods-Saxon potential using standard bound states technique code WSBETA \cite{Dudek&Nazarewicz}. Then for each state $\psi_{\nu}$ with energy $\epsilon$ we find all the  Sturmian spherical basis states (zero deformation) $\phi_{\epsilon n l j}$ with $nlj$ labels as for a spherical harmonic oscillator up to the $N_{max}$ oscillator shell. These basis states are constructed with the reaction code DWUCK4 \cite{DWUCK}. For a fixed $\epsilon$ and $nlj$ labels the code finds a scaling factor for the original spherical potential such that $\phi_{\epsilon n l j}$ is a bound state of this new potential. This scaling factor is then used to recompute the state $\phi_{\epsilon n l j}$ within the WSBETA code in the same basis where the deformed state $\psi_{\nu}$ has been computed. The expansion amplitudes 
$c_{\nu}^{nlj}$ are then calculated as described in the Appendix Section \ref{Sturmian basis expansion} and then passed back to the reaction code to compute the cross section for pickup from the deformed state $\psi_{\nu}$ using the spherical basis states $\phi_{\epsilon n l j}$. Since the $c_{\nu}^{nlj}$ amplitudes have to be added coherently, we have used the coupled channel code CHUCK3 \cite{CHUCK}.

In order to test the reliability of our amplitudes and the corresponding radial wave functions, we have compared some of our $c_{\nu}^{lj}$ and $\phi_{\epsilon l j}$ against calculations with the Rost's code 
\cite{Rost:1967}. The agreement is good but not perfect; discrepancies can be attributed to differences in the implementation of the non-zero deformation.

\begin{figure}[htb]
\centerline {\includegraphics[width=8.5cm]{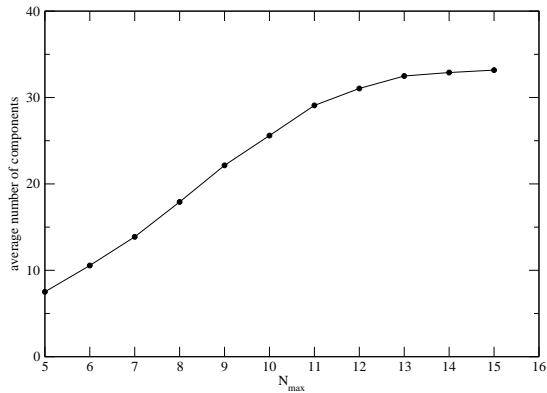}}
\caption{Model space convergence of the  Sturmian basis approach. 
The average number of components is the sum of the non-zero Sturmian 
basis components for the 47 occupied bound states divided by 47.}
\label{NNZC}
\end{figure}

In order to understand how well a state $\psi_{\nu}$ is expressed in the corresponding Sturmian basis 
one has to study the norm of $\psi_{\nu}$ in the Sturmian basis and how this 
norm converges with the size of the model space. From Fig. \ref{NNZC} one can clearly see the 
 model space convergence.

\begin{figure}[htb]
\centerline {\includegraphics[width=8.5cm]{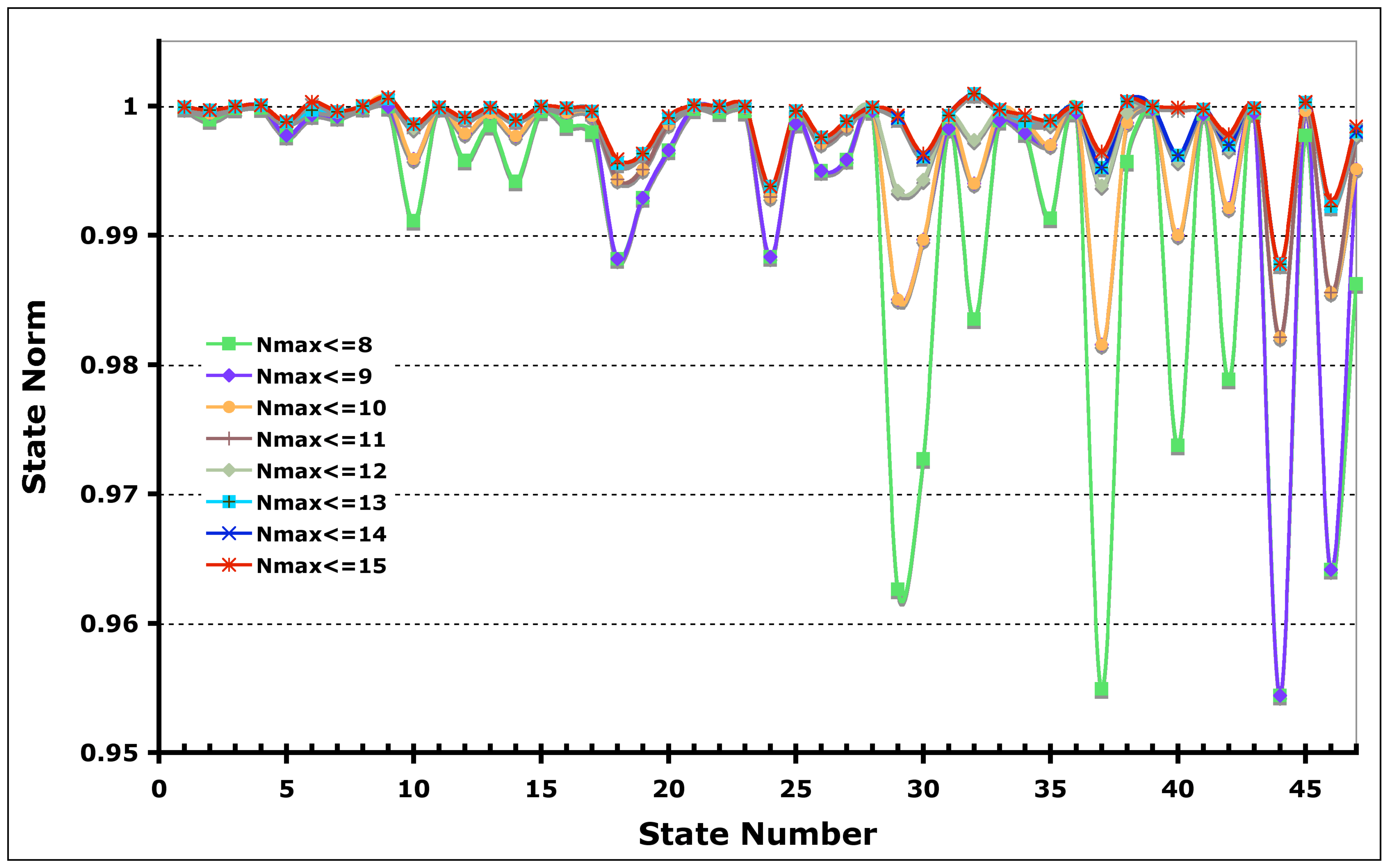}}
\caption{Norm convergence as function of the model space.}
\label{StateNorms}
\end{figure}

The state norm converges even faster.
Beyond $N_{max} =10$ the smallest non unit norm is only 0.98. Unfortunately, there are 
some particularly troubling states. As seen from Fig. \ref{StateNorms}, there are states that have 
systematically smaller norm then the other states and some are even getting norm 1.0009 due 
to numerical issues.

\begin{figure}[htb]
\centerline{\includegraphics[width=8.5cm]{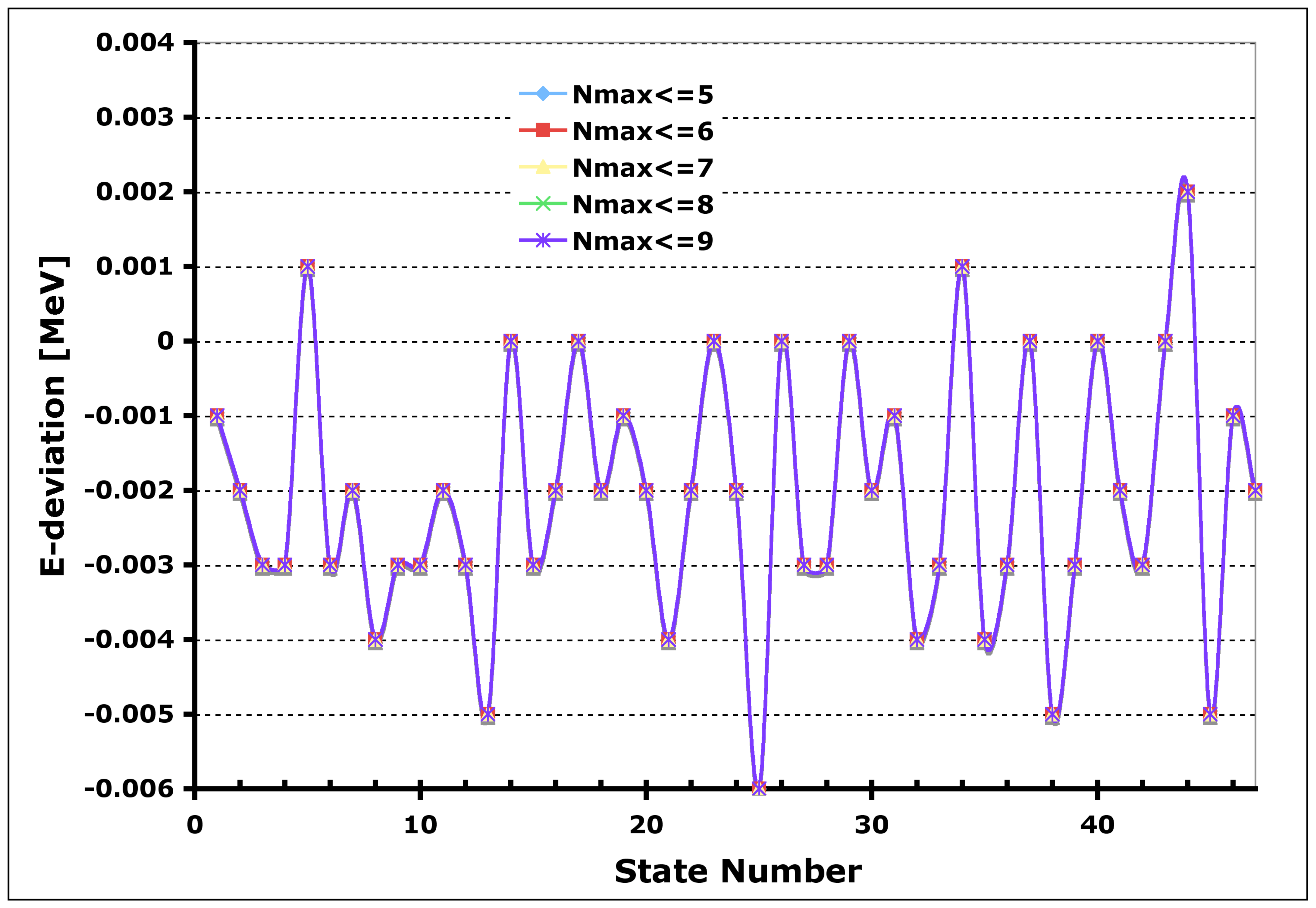}}
\centerline{\includegraphics[width=8.5cm]{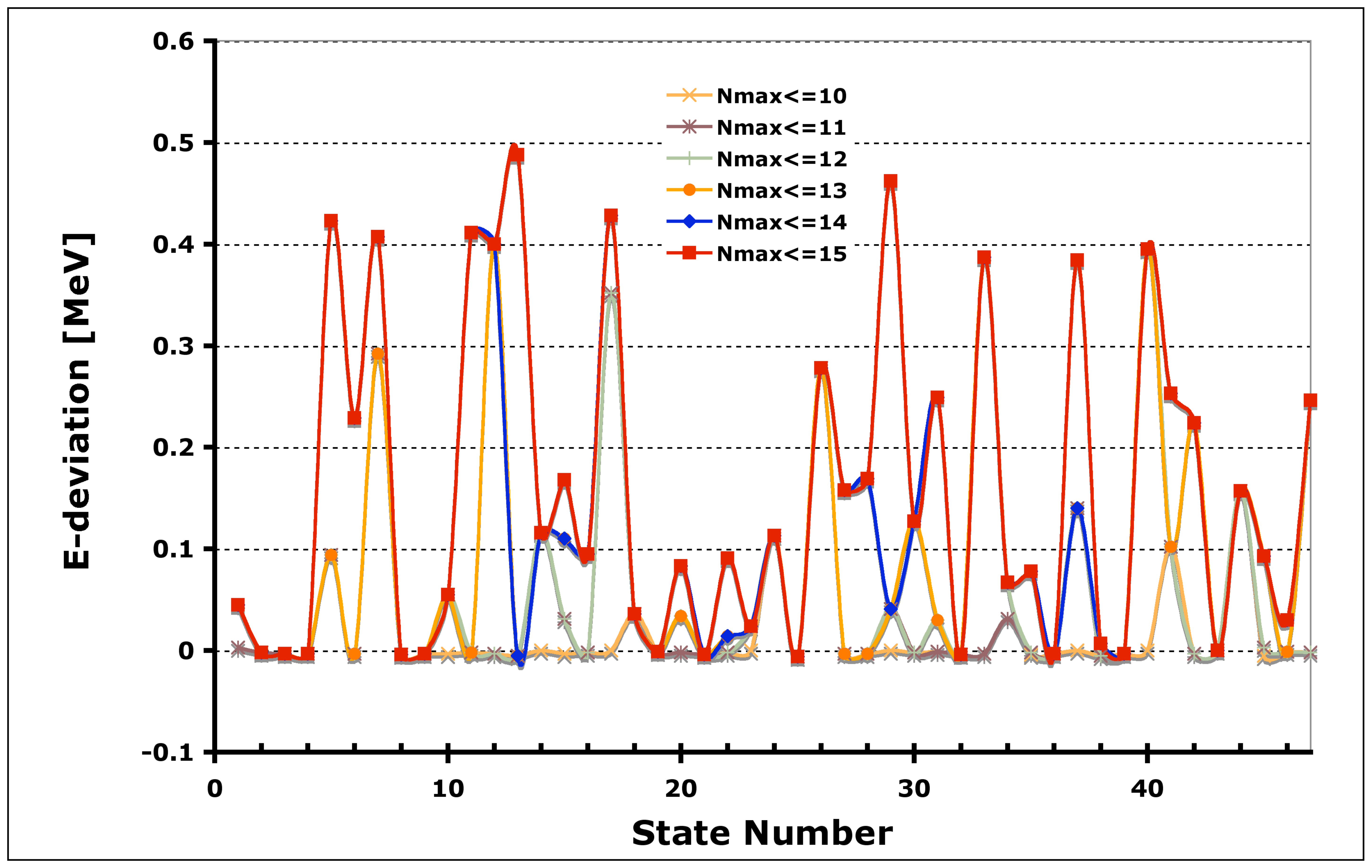}}
\caption{Maximal deviation between the actually calculated eigenenergy of a basis state and the desired energy.
Top graph: E-deviation for the model spaces $4<N_{max}<10$.
Botom graph: E-deviation for the spaces $9<N_{max}<16$.}
\label{Edeviations}
\end{figure}

It is possible that these troublesome features are actually result of a 
deeper numerical problem inherited in any standard large basis bound state technique. 
The problems are likely to be related to the fact that only the lowest states are well 
converged and thus possess the correct energy; because of that Sturmian states that 
are higher in $n$ are less accurate, which is seen from the deviation of their actual 
energy from the desired Sturmian energy shown in Fig.~\ref{Edeviations}.
As seen from the top graph in Fig. \ref{Edeviations} all basis stats computed with 
WSBETA $N_{max}<10$ have deviation within few keV. However, for larger model spaces  
$N_{max}>9$ there are basis stats whose deviation is getting much too big. This could 
become one of the main issues for the presented computational method.

\subsection{Neutron pickup cross sections}

\begin{figure}[htb]
\centerline {\includegraphics[width=9.5cm]{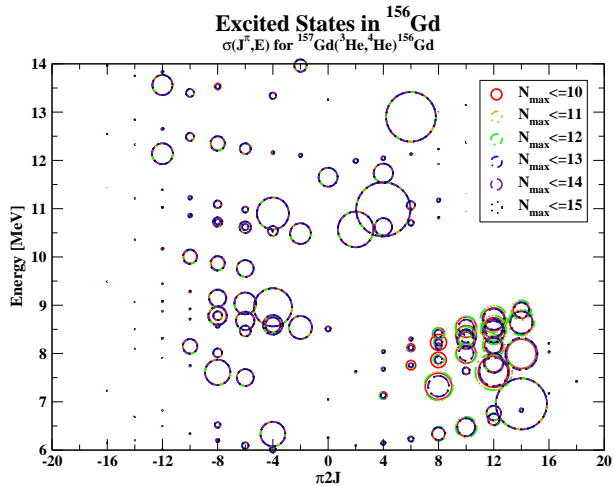}}
\caption{Individual cross sections for neutron transfer from a deformed state that results in a 
final state with $J^{\pi}$ and energy $E$ as shown in Fig.~\ref{pi2JE}. The size of the circles 
reflects the size of the corresponding cross sections.}
\label{EpiJ2sigma}
\end{figure}

The individual cross sections, for neutron transfer from a deformed state $\psi_{\nu}$ that results in a 
final state with $J^{\pi}$ and energy $E$ as shown in Fig.~\ref{pi2JE}, are calculated as coherent 
cross sections with the code CHUCK3 \cite{CHUCK} using the amplitudes $c^{\nu}_{nlj}$ times a 
Clebsch-Gordan coefficient and other appropriate factors  \cite{Andersen&Back:1970}:
\begin{equation}
\sqrt{\frac{(1+\delta_{0,K_{i}K_{f}})}{2j+1}}D_{0}\times c^{\nu}_{nlj}\times(J_{f}K_{f}| j m, J_{i},K_{i}).
\end{equation}
Here $D_{0}$  is the strength of the zero range transfer potential 
($D_{0}\delta (x)$, $D_{0}^{2}$=18 \cite{DWUCK}).
The resulting reaction cross sections defined for each single particle state 
$\sigma_{\lambda}(\epsilon_{\nu},J^{\pi})$ are shown in Fig.~\ref{EpiJ2sigma}.
In what follows the index $\lambda$ will denote the equivalent pair of indexes $(\epsilon_{\nu},J^{\pi})$.
The size of the circles reflects the size of the corresponding cross sections.
As seen from Fig.~\ref{EpiJ2sigma} there is often a good overlap of the circles 
for the model spaces shown. This indicates the convergence of the calculated cross sections.

The cross sections that one can calculate within the presented framework correspond to 
sharp final states. In reality there are widths associated with the single particle states as well 
as with the final states. In order to produce a smooth total cross section as a function of the 
excitation energy of the $^{156}$Gd system we consider a Cauchy type smearing 
distribution function  (Lorentzian) and define a smooth $\sigma(E)$ cross section \cite{Andersen&Back:1970}:
\begin{eqnarray}
\rho_{\lambda}(E)&=&\frac{1}{2\pi}\frac{4\Gamma}{4(E-E_{\nu})^{2}+\Gamma^{2}}\\
\sigma(E)&=&\sum_{\lambda}\rho_{\lambda}(E)\sigma_{\lambda}, ~\Gamma=a+b E. \nonumber
\end{eqnarray}

\begin{figure}[htb]
\centerline {\includegraphics[width=8.5cm]{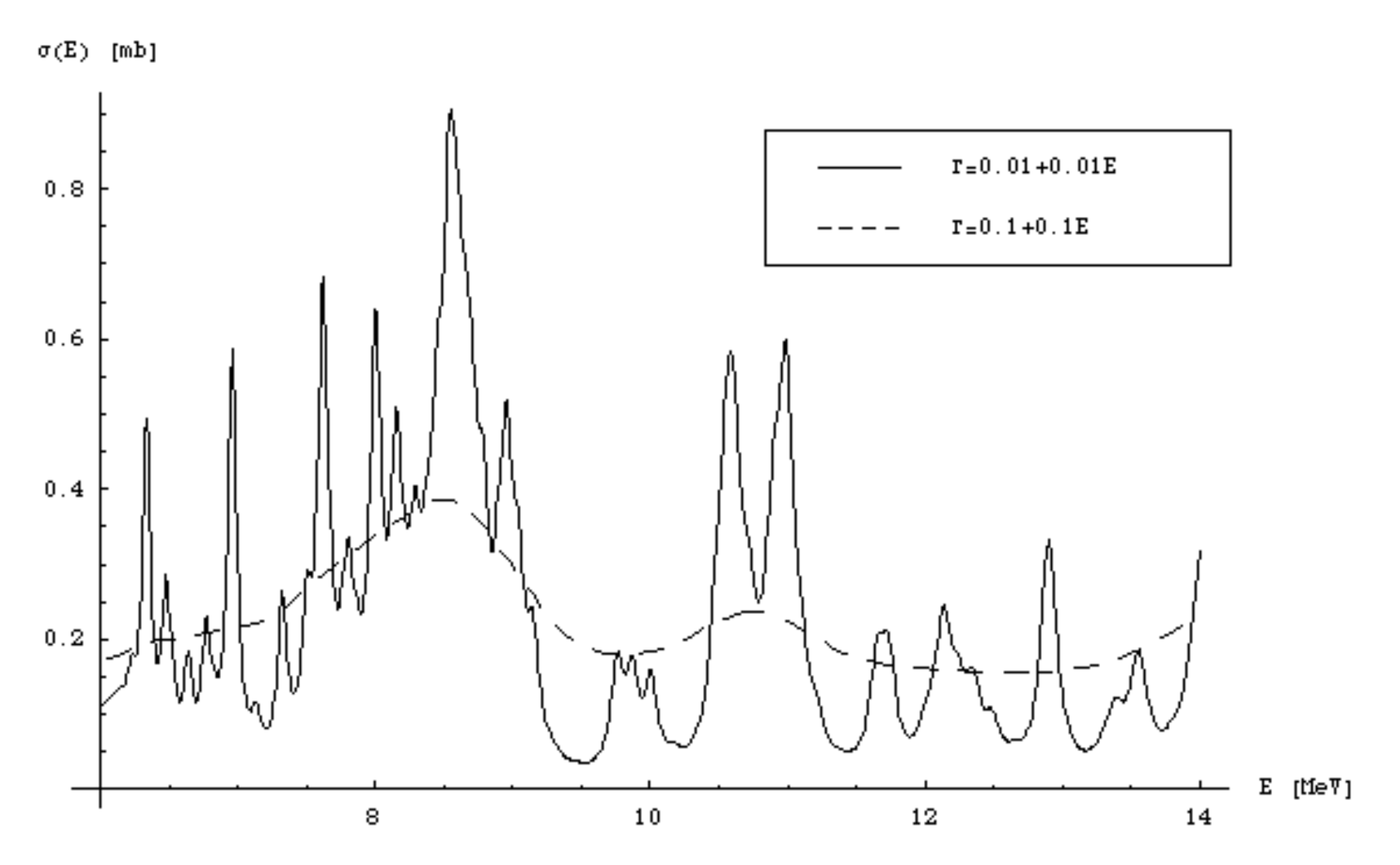}}
\caption{Total neutron cross section $\sigma(E)$ in mb units smeared with Lorentzian using 
energy dependent width $\Gamma=a+b E$. Dashed line represents $\Gamma=0.1+0.1E$ and 
continuous line represents $\Gamma=0.01+0.01E$.}
\label{Smearing}
\end{figure}

In Fig.\ref{Smearing} are shown the effects on the total cross section $\sigma(E)$ of different 
smearing width choices $\Gamma$. The choice of $\Gamma$ should eventually be fixed by 
comparison of the theoretical and experimental cross sections. 

\begin{figure}[htb]
\centerline {\includegraphics[width=8.5cm]{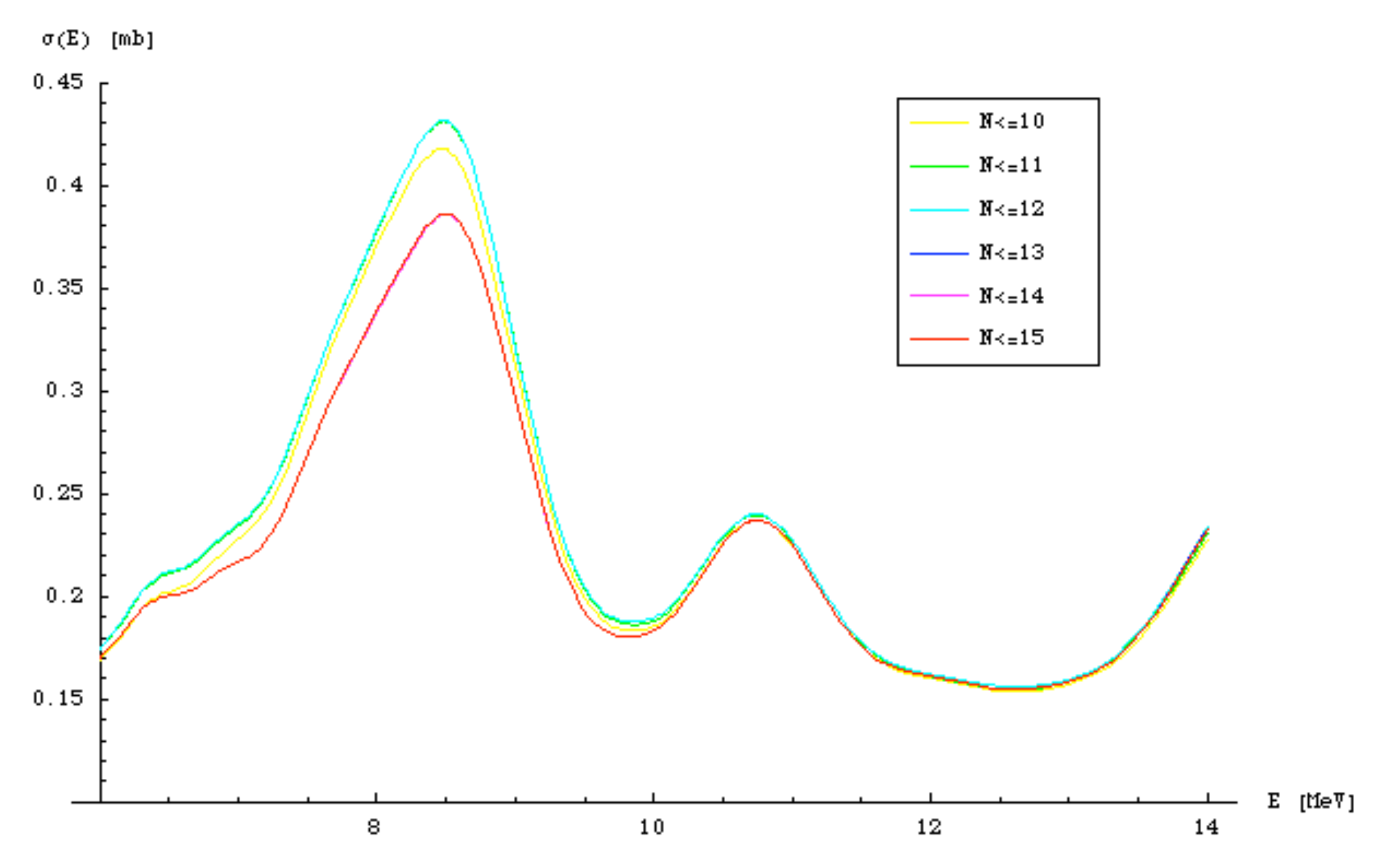}}
\caption{Smeared total cross section for $\Gamma=0.1+0.1E$ as a function of $N_{max}$. 
The identical $\sigma(E)$  for $N_{max}>12$ (red curve) illustrate the convergence of the results.}
\label{Convergence}
\end{figure}

For each choice of a model space $N_{max}$ one can compute a set of cross sections 
$\sigma_{\lambda}(\epsilon_{\nu},J^{\pi})$. If the model space is sufficiently big so that 
convergence has been archived for the  $c_{\nu}^{nlj}$ then one expects that the corresponding 
cross sections would be the same for the converged spaces. It was show in Fig. \ref{NNZC} 
that convergence requires $N_{max}>12$. Therefore, as expected, the smeared total cross 
sections for $N_{max}>12$ are the same as seen in Fig. \ref{Convergence}.


\subsection{The $P(J^{\pi},E)$ distributions}

The smeared cross sections, introduced in the previous section, can be used to define 
the probability to excite a state with quantum numbers $J^{\pi}$:
\begin{eqnarray}
P(J^{\pi};E)=\frac{1}{\sigma(E)}\sum_{\lambda}
\delta_{J,J_{\lambda}}
\delta_{\pi,\pi_{\lambda}}
\rho_{\lambda}(E)\sigma_{\lambda}
\end{eqnarray}
It is clear that if one sums over all possible $J^{\pi}$  quantum numbers then one gets one,
that is, $\sum_{J^{\pi}}P(J^{\pi};E)=1$.

\begin{figure}[htb]
\centerline {\includegraphics[width=9cm]{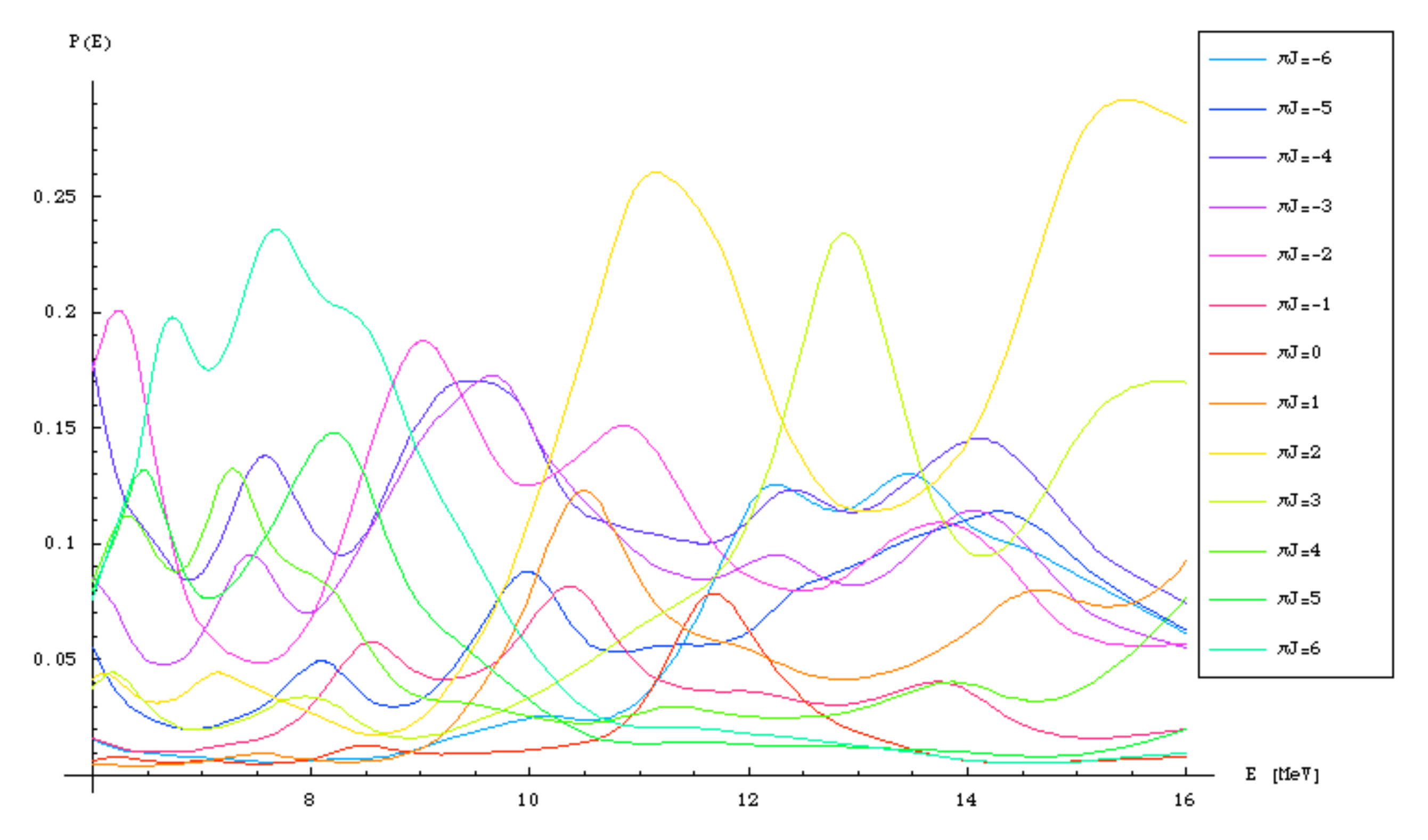}}
\caption{All $P(J^{\pi};E)$ distributions for $J<8$.}
\label{JpiE_dist}
\end{figure}

In Fig. \ref{JpiE_dist} are shown the formal probabilities $P(J^{\pi};E)$ for exciting 
$J^{\pi}$ state of $^{156}$Gd at energy $E$ through direct neutron pickup via 42 MeV $^{3}$He
on $^{157}$Gd target.

\begin{figure}[htb]
\centerline{\includegraphics[width=8cm]{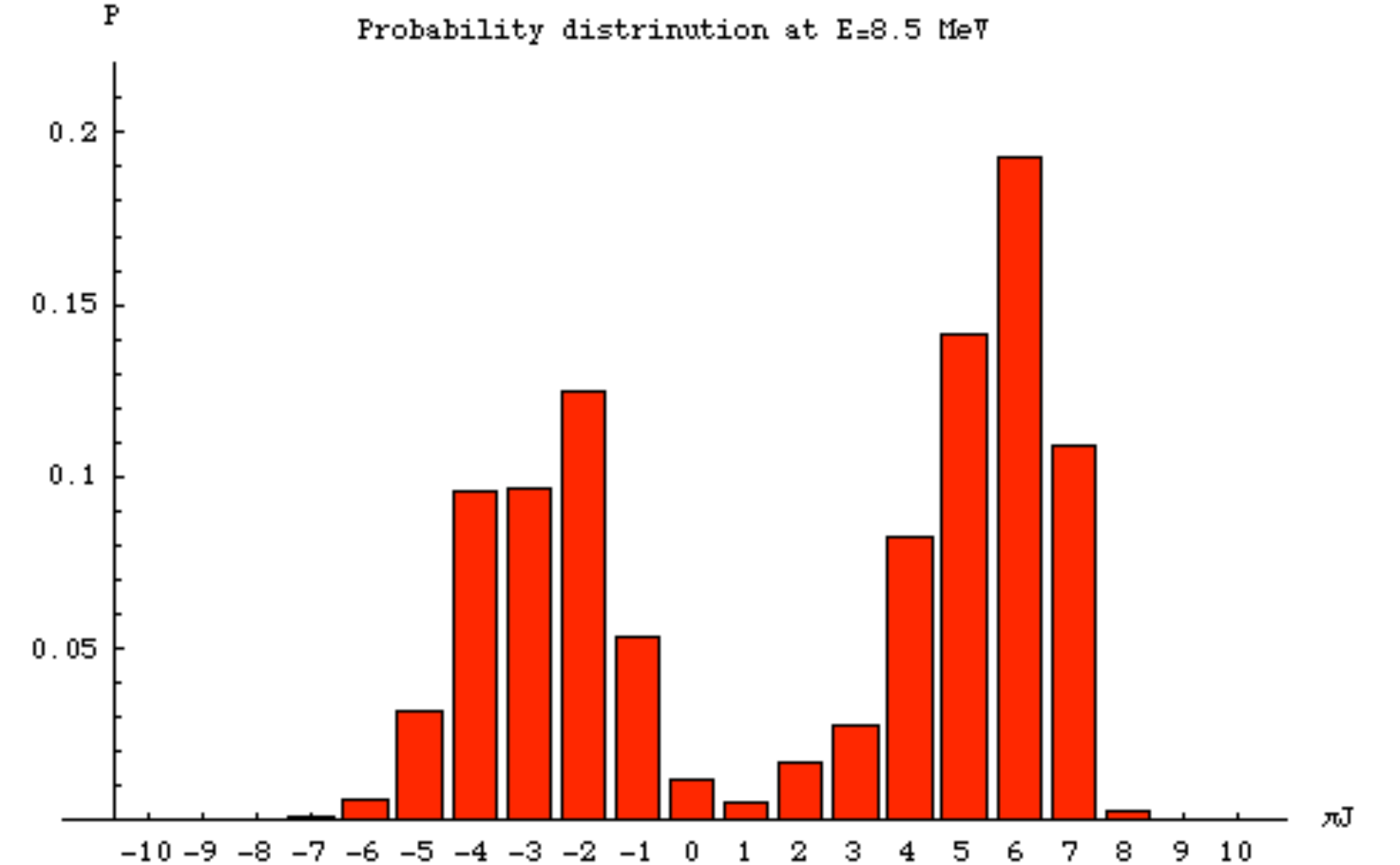}}
\centerline{\includegraphics[width=8cm]{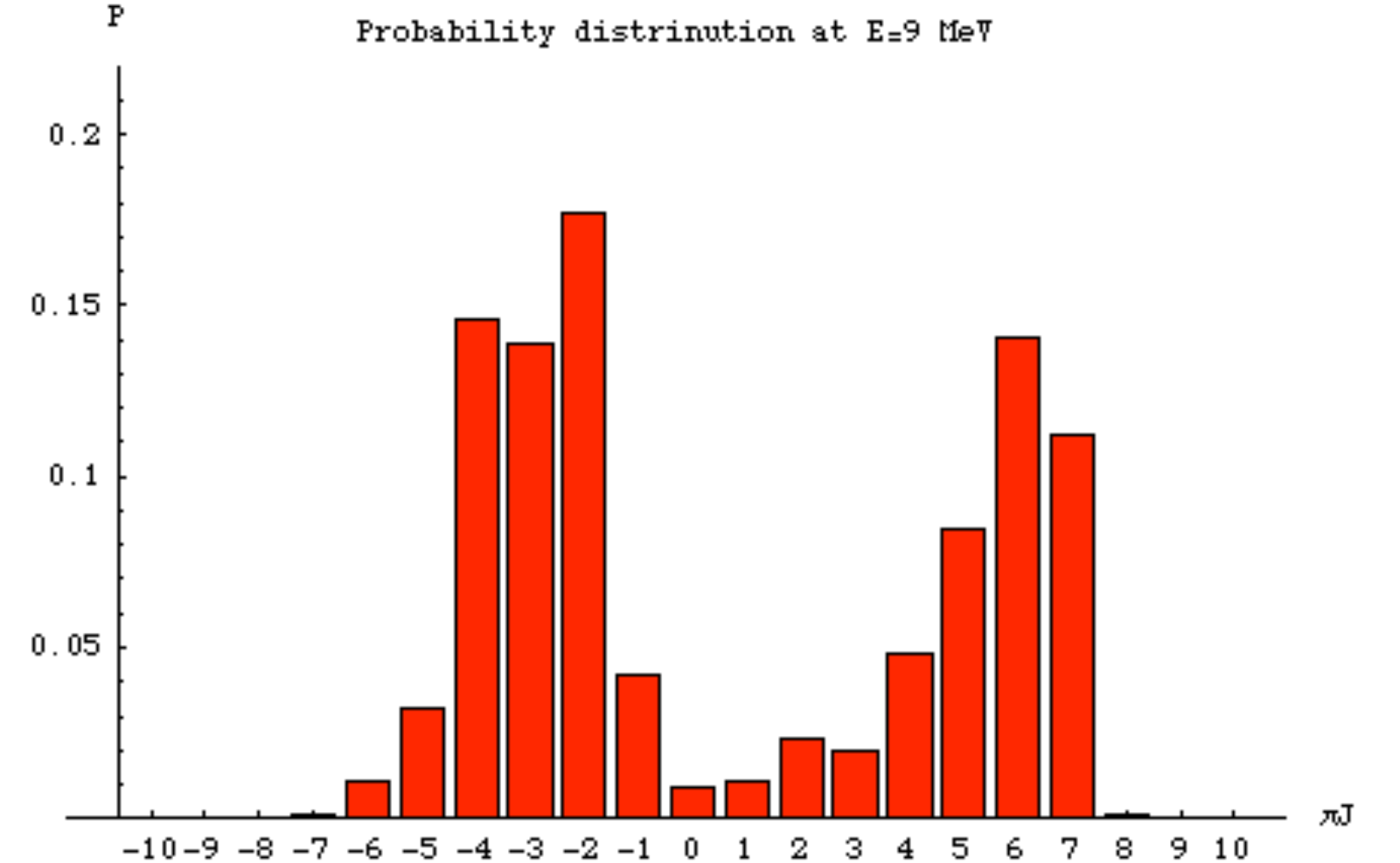}}
\centerline{\includegraphics[width=8cm]{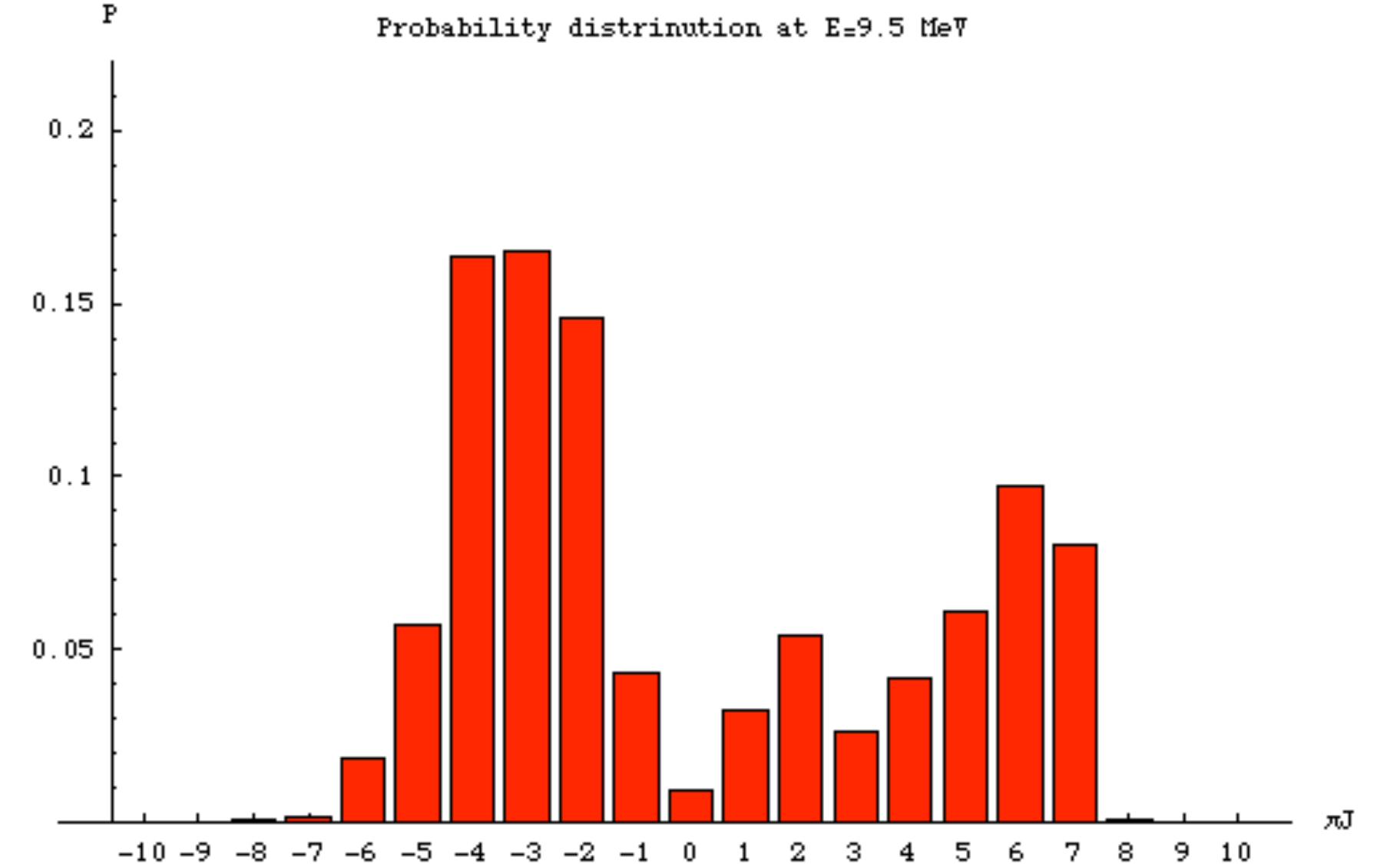}}
\caption{$P(J^{\pi};E)$ distributions for energies near the neutron separation energy 
in $^{156}$Gd. From top to bottom the energies are  8.5, 9, and 9.5 MeV. 
The sign of the horizontal coordinate corresponds to the parity $\pi$ and its magnitude gives $J$.}
\label{3Jpi_dist}
\end{figure}

In Fig. \ref{3Jpi_dist} are shown $P(J^{\pi};E)$ distributions that are of interest to the surrogate method. 
These graphs show that at least in principle there are no dead channels due to zero  $P(J^{\pi})$ and thus any channel that goes through $J < 8$ should be able to provide information on the decay probability of the 
compound system. The apparently zero values of the distribution for $J>6$ are due to the model space 
used that focuses mostly on $J<8$ excited states in $^{156}$Gd.

\section{Concluding remarks and discussions}

The present computational method is clearly a viable approach to particle pick-up reactions in 
deformed  and  strongly deformed systems. However, for stripping reactions the method would 
need to incorporate the effects of the pairing interaction since adding a nucleon would involve 
mostly levels near and above the Fermi level and even unbound states and resonances. 
In heavier nuclei the treatment of the valence single particles states above the Fermi level would 
probably face a problem due to the treatment of weakly bound states that may render the current 
computational  technique useless; this could easily be monitored through evaluation of 
the norm of the bound states $\psi_{\varepsilon m}$ and the quality of the Sturmian basis 
$\phi_{\varepsilon nljm}$. An alternative to the current approach that uses a standard bound state 
computational technique to compute the overlap of the state $\psi_{\varepsilon m}$ and 
$\phi_{\varepsilon nljm}$ would be to extend Rost's code to include higher multipoles beyond 
the $\beta_{2}$ or to solve numerically for $\psi_{\varepsilon m}$ as generalized eigenvalue 
problem for non zero deformation using the Sturmian basis states $\phi_{\varepsilon nljm}$.


The surrogate method (Fig. \ref{TheSurrogateMethod}) assumes that the surrogate reaction populates 
$J^{\pi}$ states of the  intermediate nuclear system within the same energy range as the desired reaction.
The method would certainly fail if the distribution $P(J^{\pi},E)$ becomes zero for some relevant values of 
$J^{\pi}$ at the relevant range of energies $E$ since there would be no way of deducing the decay 
probabilities $g_{\chi}(J^{\pi},E)$. If $P(J^{\pi},E)=0$ for a given $J^{\pi}$ at an isolated energy $E$ one could perhaps devise an analytic continuation for $g_{\chi}(J^{\pi},E)$. Our calculations, however, show that 
within the assumptions and computational modeling the reaction 
$^{3}$He+$^{157}$Gd $\rightarrow$ $^{4}$He+$^{156}$Gd$^{\star}$ 
has a well behaved formation probability $P(J^{\pi},E)$ (see Fig.~\ref{JpiE_dist} and Fig.~\ref{3Jpi_dist}) 
within a wide energy range relevant to the desired reaction 
$^{155}$Gd+n $\rightarrow$ $^{156}$Gd$^{\star}$.
Therefore, given an experimental input on the decay probability $P_{\delta\chi}$ into an exit 
channel $\chi$ within the surrogate formation channel $\delta$, one should in principle be 
able to determine $g_{\chi}(J^{\pi},E)$.

Experimental input is essential for the fine tuning of the model parameters such as single 
particle Woods-Saxon potential,  optical model potential for the incoming and out going projectiles 
as well as the choice of $\Gamma(E)$ used in the smearing function $\rho(E)$. Above all a comparison 
with experiment should be the true measure of the applicability the surrogate method.

\acknowledgments{
We would like to thank  P. Navratil, J. P. Vary, and W. Younes  for interesting and helpful discussions.
This work was partly performed under the auspices of the U. S. Department of Energy 
by the University of California, Lawrence Livermore National Laboratory under contract 
No. W-7405-Eng-48. Support was also provided from the LDRD contract No.~04--ERD--057.
Some of the calculations have been performed using the LLNLÕs Thunder machine.}

\section{Appendix}
\subsection{$\Omega_{\pm}$ splitting due to particle-core coupling}
\label{Omega_pm splitting}
A more rigorous analysis  in the intrinsic frame of the rotational energy of the hole in the 
core system gives:
\begin{eqnarray}
&&<\Omega_{\pm}|J\cdot T\cdot J|\Omega_{\pm}>=\nonumber\\
~&&=2H_{C}-\frac{\hbar^{2}}{2{\cal{I}}_{\perp}}(<K|I^{2}|K>+<\nu | j^{2}|\nu>)\nonumber\\
~&&-\frac{\hbar^{2}}{2{\cal{I}}_{\perp}}
\left[
{\left( \frac{{\cal{I}}_{\perp}}{{\cal{I}}_{\parallel}}-1\right)}(K^{2}+\nu^{2}) 
\pm 2\frac{{\cal{I}}_{\perp}}{{\cal{I}}_{\parallel}}K\nu
\right]
\end{eqnarray}
Here $\Omega_{\pm}=|K\pm\nu|$, $J$ is the intrinsic angular momentum of the 
combined system (hole and core),  $I$ for the core, and $j$  for the hole.
$H_{C}$ is the Coriolis coupling (\ref{Coriolis coupling}).
$T$ is the moment of inertia of the combined system. 
$T$ is diagonal in the intrinsic frame with $T_{x}=T_{y}=1/(2 {\cal I}_{\perp})$ and
$T_{z}=1/(2 {\cal I}_{\parallel})$.

For the shape and moment of inertia of rigid ellipsoid of mass $M$ one has:
\begin{eqnarray}
&& {\cal I}_{i}=\frac{M}{5}\sum_{j}d_{j}^{2}-d_{i}^{2},~R(\theta,\phi)=R_{0}(1+\beta Y_{20}(\theta,\phi))\nonumber \\
&&d_{x}=d_{y}=1- \beta\frac{1}{4}\sqrt{\frac{5}{\pi }} \neq d_{z}=1+ \beta \frac{1}{2} \sqrt{\frac{5}{\pi }}
\end{eqnarray}
Which gives:
\begin{equation}
\frac{{\cal{I}}_{\perp}}{{\cal{I}}_{\parallel}}=
\frac{32 \pi+\beta  \left(25 \beta +8 \sqrt{5 \pi }\right) }{2 \left(\sqrt{5} \beta -4 \sqrt{\pi }\right)^2}
\end{equation}

\subsection{$c_{\nu}^{nlj}$ amplitudes in a Sturmian basis}
\label{Sturmian basis expansion}

As we already mentioned, for strong deformation one needs to consider either
the full Taylor expansion \cite{Glendenning:2004} or to use numerical
methods to solve the Schr\"{o}dinger equation \cite{Dudek&Nazarewicz}. Given
the code WSBETA \cite{Dudek&Nazarewicz} along with the code DWUCK \footnote{
DWUCK or any other code that can provide search on the potential depth for a
fixed binding energy is needed to determine the potential strength needed to
place the $nlj$ spherical state at the desired binding energy.} plus a few 
post-processing codes one can solve the Sturmian problem and find 
expansion of any deformed state $\psi_{\varepsilon m }$ 
($\beta \neq 0$) in terms of non-orthogonal Sturmian
states $\phi _{\varepsilon nljm}$ with correct tails given by their binding
energy $\varepsilon $. 
\begin{equation}
\psi _{\varepsilon m}=\sum_{nlj}c^{nlj}\phi _{\varepsilon nljm}
\label{def wf in sph basis}
\end{equation}
For this purpose, one solves the following set of eigenvalue equations: 
\begin{eqnarray}
H_{\beta \neq 0}\psi _{\varepsilon \Omega } &=&\varepsilon \psi
_{\varepsilon \Omega }  \label{tail energy condition} \\
H_{\beta =0}\phi _{\varepsilon nljm} &=&\varepsilon \phi _{\varepsilon nljm}
\nonumber
\end{eqnarray}
Since WSBETA gives solutions to these problems in the same cylindrical
harmonic oscillator basis $\left| NN_{z}\Lambda \Omega \right\rangle ,$ one
can evaluate the overlap matrix of these states: 
\begin{eqnarray*}
\phi _{\varepsilon nljm} &=&\sum_{NN_{z}\Lambda }D_{nljm}^{NN_{z}\Lambda
}\left| NN_{z}\Lambda m\right\rangle  \\
\psi _{\varepsilon \Omega } &=&\sum_{NN_{z}\Lambda }B_{\Omega
}^{NN_{z}\Lambda }\left| NN_{z}\Lambda \Omega \right\rangle 
\end{eqnarray*}
Here the quantum numbers $N,N_{z},\Lambda ,$ and $m$ represent the total
number of oscillator quanta, the number of oscillator quanta in the $z$
-direction, and the projection of the angular momentum and the total angular
momentum onto the symmetry axis $z$. Since Sturmian states $\phi
_{\varepsilon nljm}$ with different $ljm$ quantum numbers are orthonormal due
to orthogonality of the spin and angular momentum components, we have
non-orthogonality only within a fixed $ljm$ sub-set: 
\begin{equation}
\mu _{nk}^{ljm}=\left\langle \phi _{\varepsilon nljm}|\phi _{\varepsilon
kljm}\right\rangle =\sum_{NN_{z}\Lambda }\bar{D}_{nljm}^{NN_{z}\Lambda
}D_{kljm}^{NN_{z}\Lambda }  \label{overlap mu}
\end{equation}
\begin{equation}
\Xi _{nljm}=\left\langle \phi _{\varepsilon nljm}|\psi _{\varepsilon
m}\right\rangle =\sum_{NN_{z}\Lambda }\bar{D}_{nljm}^{NN_{z}\Lambda
}B_{m}^{NN_{z}\Lambda }  \label{dot product}
\end{equation}
In order to determine the coefficients $c^{nlj}$ in (\ref{def wf in sph
basis}) we have to use the inverse $\mu _{ljm}^{nk}$ of the overlap $\mu
_{nk}^{ljm}$ ($\sum_{k}\mu _{nk}\mu ^{n^{\prime }k}=\delta _{n}^{n^{\prime }}
$). Within each sub-space of fixed $lj$ we have: 
\[
\Xi _{n}=\sum_{k}c^{k}\mu _{nk}{\rm ~-fixed~}lj\Rightarrow c^{n}=\sum_{k}\mu
^{nk}\Xi _{k}
\]
In general one expects to have no problems in inverting $\mu _{nk}$ since 
$\det \mu =0$ would imply liner dependence of the vectors $\phi _{\varepsilon
nljm}$ which should not be present since each of these states has different
number of nodes. How many Sturmian states $\phi _{\varepsilon nlj}$ are
needed is determined by the normalization condition: 
\[
\left\| \psi _{\varepsilon \Omega }\right\| ^{2}=1\geq \sum_{lj}\sum_{nk}
\bar{c}^{n}\mu _{nk}c^{k}=\sum_{lj}\sum_{n}\bar{c}^{n}\Xi _{n}
\]
Therefore the number $N_{S}$ of Sturmian states $\phi _{\varepsilon nlj}$ is
determined by the normalization requirement: 
\begin{equation}
\sum_{lj}^{j_{\max }}\sum_{n,k}^{N_{S}}\bar{\Xi}_{n}\mu ^{nk}\Xi _{k}=1
\label{normalization condition}
\end{equation}
Here an implicit $lj$ index is assumed for ${\Xi }_{n}$ and $\mu ^{nk}$.
Since parity considerations determine the value of $l$ for any given $j$ one
can write all $lj$ indexed partitions as single $j$ index.

When the asymptotic tail is proper (\ref{tail energy condition}) and $N_{S}$
and $j_{\max }$ are such that the normalization condition (\ref
{normalization condition}) is satisfied then one can be sure that there are
no states omitted, continuum or discrete, because we have normalized bound
state with the correct expansion at small $r$ as well as at large $r$.

\subsection{Phase conventions for the wave functions}

The overall phase of a wave function, single particle wave function in particular, 
is arbitrary and unobservable. Thus, it is usually fixed for convenience by a suitable choice 
of convention. Although there could be as many conventions as wave function, there 
are two main choices in the literature. One convention is to consider wave functions that 
are positive as $r\rightarrow \infty$ \cite{DWUCK,CHUCK}. Another convention is to consider 
wave functions that are positive at the origin $r\rightarrow 0$ \cite{Hird&Huang:CPC1975}.
Sometime the convention is not clear at all \cite{Dudek&Nazarewicz}.

For our calculations, it is important to adjust the $c^{\nu}_{nlj}$ since we use two different codes 
\cite{Dudek&Nazarewicz, CHUCK}. First the $c^{\nu}_{nlj}$ are calculated from the WSBETA 
\cite{Dudek&Nazarewicz} wave functions as described in Appendix section 
\ref{Sturmian basis expansion}. Then we take into account the phase difference relative to 
the wave functions generated by DWUCK4 and CHUCK3 \cite{DWUCK,CHUCK}.
For this purpose, we evaluate the sign of the WSBETA spherical wave functions that are given in 
cylindrical basis at $\theta=\pi/3, \phi=0$, and $r\rightarrow \infty$ using an asymptotic expression 
of the cylindrical basis wave functions.

\bibliographystyle{apsrev}
\bibliography{RCSforDN}

\end{document}